\documentclass[%
reprint,
superscriptaddress,
showpacs,
preprintnumbers,
 bibnotes,
 amsmath,amssymb,
 aps,
prl
]{revtex4-1}

\usepackage{graphicx}
\usepackage{dcolumn}
\usepackage{bm}
\usepackage{color}

\begin{document}


\title{Raman scattering study of tetragonal magnetic phase in {Sr$_{1-x}$Na$_x$Fe$_2$As$_2$}: structural symmetry and electronic gap}

\author{Li~Yue}
\affiliation{International Center for Quantum Materials, School of Physics, Peking University, Beijing 100871, China}
\author{Xiao~Ren}
\affiliation{International Center for Quantum Materials, School of Physics, Peking University, Beijing 100871, China}
\author{Tingting~Han}
\affiliation{International Center for Quantum Materials, School of Physics, Peking University, Beijing 100871, China}
\author{Jianqing~Guo}
\affiliation{International Center for Quantum Materials, School of Physics, Peking University, Beijing 100871, China}
\author{Zhicheng~Wu}
\affiliation{International Center for Quantum Materials, School of Physics, Peking University, Beijing 100871, China}
\author{Yan~Zhang}
\affiliation{International Center for Quantum Materials, School of Physics, Peking University, Beijing 100871, China}
\affiliation{Collaborative Innovation Center of Quantum Matter, Beijing 100871, China}
\author{Yuan~Li}
\email[]{yuan.li@pku.edu.cn}
\affiliation{International Center for Quantum Materials, School of Physics, Peking University, Beijing 100871, China}
\affiliation{Collaborative Innovation Center of Quantum Matter, Beijing 100871, China}

\begin{abstract}
We use inelastic light scattering to study Sr$_{1-x}$Na$_x$Fe$_2$As$_2$ ($x\approx0.34$), which exhibits a robust tetragonal magnetic phase that restores the four-fold rotation symmetry inside the orthorhombic magnetic phase. With cooling, we observe splitting and recombination of an $E_g$ phonon peak upon entering the orthorhombic and tetragonal magnetic phases, respectively, consistent with the reentrant phase behavior. Our electronic Raman data reveal a pronounced feature that is clearly associated with the tetragonal magnetic phase, suggesting the opening of an electronic gap. No phonon back-folding behavior can be detected above the noise level, which implies that any lattice translation symmetry breaking in the tetragonal magnetic phase must be very weak.

\end{abstract}


\maketitle

The iron-based superconductors exhibit rich phase diagrams due to the interplay among charge, spin, orbital, and lattice degrees of freedom \cite{StewartRevModPhys2011,PaglioneNatPhys2010,JohnstonAdvPhys2010}. One of the key features in the phase diagram of the pnictides is the closely-related magnetic and nematic transitions, which break the O(3) spin rotation symmetry and lower the lattice four-fold rotation symmetry C$_4$ down to C$_2$, respectively, resulting in an orthorhombic spin-density-wave ($o$-SDW) phase with stripe-like staggered in-plane magnetic moments. The microscopic origin of this magneto-nematic transition has aroused intense research interest. While the transition is widely considered to be electronic, both orbital \cite{KontaniPRB2011} and spin degrees of freedom have been proposed as the driving force \cite{FernandesNatPhys2014}, and the spin scenarios can be sub-divided into local-moment models based on exchange interactions \cite{FangPRB2008,KrugerPRB2009,XuPRB2008} and itinerant models based on Fermi-surface nesting \cite{LorenzanaPRL2008,FernandesPRB2012}. Very recent experiments further suggest that the strength of spin-orbit interactions is non-negligible \cite{WatsonPRB2015,JohnsonPRL2015,BorisenkoNatPhys2016,MaPreprint2016}, so that the spin and the orbital degrees of freedom might also cooperate \cite{CvetkovicPRB2013,ChristensenPRB2015}.

A pivotal fact in support of the spin-nematic scenario is that a novel tetragonal SDW ($t$-SDW) phase can take over inside the $o$-SDW phase and, upon the formation of the new magnetic order, the C$_4$ lattice symmetry is restored.
First discovered in Ba$_{1-x}$Na$_x$Fe$_2$As$_2$ \cite{AvciNatCommu2014} and later universally found in hole-doped ``122'' iron pnictides \cite{AllredPRB2015,TaddeiPRB2016,TaddeiPRB2017}, the $t$-SDW phase is characterized by its double-$\mathbf{Q}$  magnetic structure, which can be understood as a superposition of two SDW phases with perpendicular wave vectors (0,$\pi$) and ($\pi$,0).
Moreover, the $t$-SDW transition features a distinct spin reorientation from the $ab$ plane (in the $o$-SDW phase) to the $c$ axis \cite{WasserPRB2015,AllredPRB2015,MallettEPL2015,AllredNatPhys2016}.
Apart from restoring the C$_4$ symmetry, the $t$-SDW phase is very important for understanding the iron-based superconductors in more general contexts:
(1) From the double-$\mathbf{Q}$ and collinear magnetic structure, it is expected that half of the iron sites have vanished magnetic moments and the other half have doubled moments, as has been confirmed experimentally \cite{MallettEPL2015,AllredNatPhys2016}. This in turn favors description of the magnetism from an itinerant-electron standpoint \cite{AllredNatPhys2016}.
(2) The competition between the $o$-SDW and the $t$-SDW phases underscores the importance of spin-orbit interactions \cite{CvetkovicPRB2013,ChristensenPRB2015}, as the $c$-polarization of the ordered moments in the $t$-SDW phase resembles spin-space anisotropy in the low-energy magnetic excitations in the nematic phases of the ``122'' pnictides \cite{QureshiPhysRevB2012,WangCPhysRevX2013,SteffensPhysRevLett2013,ZhangPhysRevB2013,Wasser2016} and FeSe \cite{MaPreprint2016}. (3) Superconductivity competes with the $t$-SDW phase more strongly than with the $o$-SDW phase \cite{AvciNatCommu2014,MallettPRL2015,BohmerNatCommu2015,GastiasoroPRB2015}. Upon cooling from the $t$-SDW phase into the superconducting phase, the lattice symmetry may even go back to C$_2$ \cite{MallettPRL2015,BohmerNatCommu2015,LiuCryst2016}.

Here we report a Raman scattering study of the sequential C$_4$-C$_2$-C$_4$ transitions in Sr$_{1-x}$Na$_x$Fe$_2$As$_2$. We first confirm that these transitions leave clear signatures in the energy spectrum of Brillouin zone-center phonons, consistent with the expected changes in the point-group symmetry. Upon cooling into the $t$-SDW phase, we observe a pile-up of intensity in the electronic Raman spectrum near 260 cm$^{-1}$ along with a depletion at lower energies, suggesting the opening of an electronic gap of similar sizes. While electronic band reconstructions might be caused by charge-density modulations \cite{ChristensenPRB2015,GastiasoroPRB2015,FernandesPRB2016} that concurrently develop with the $t$-SDW order, our deliberate search for back-folding behaviors of phonons allows us to place a tight upper bound on the intensity of the folded branches, implying that the breaking of lattice translation symmetry, if any, must be very weak.

\begin{figure}[t]
\includegraphics[width=3in]{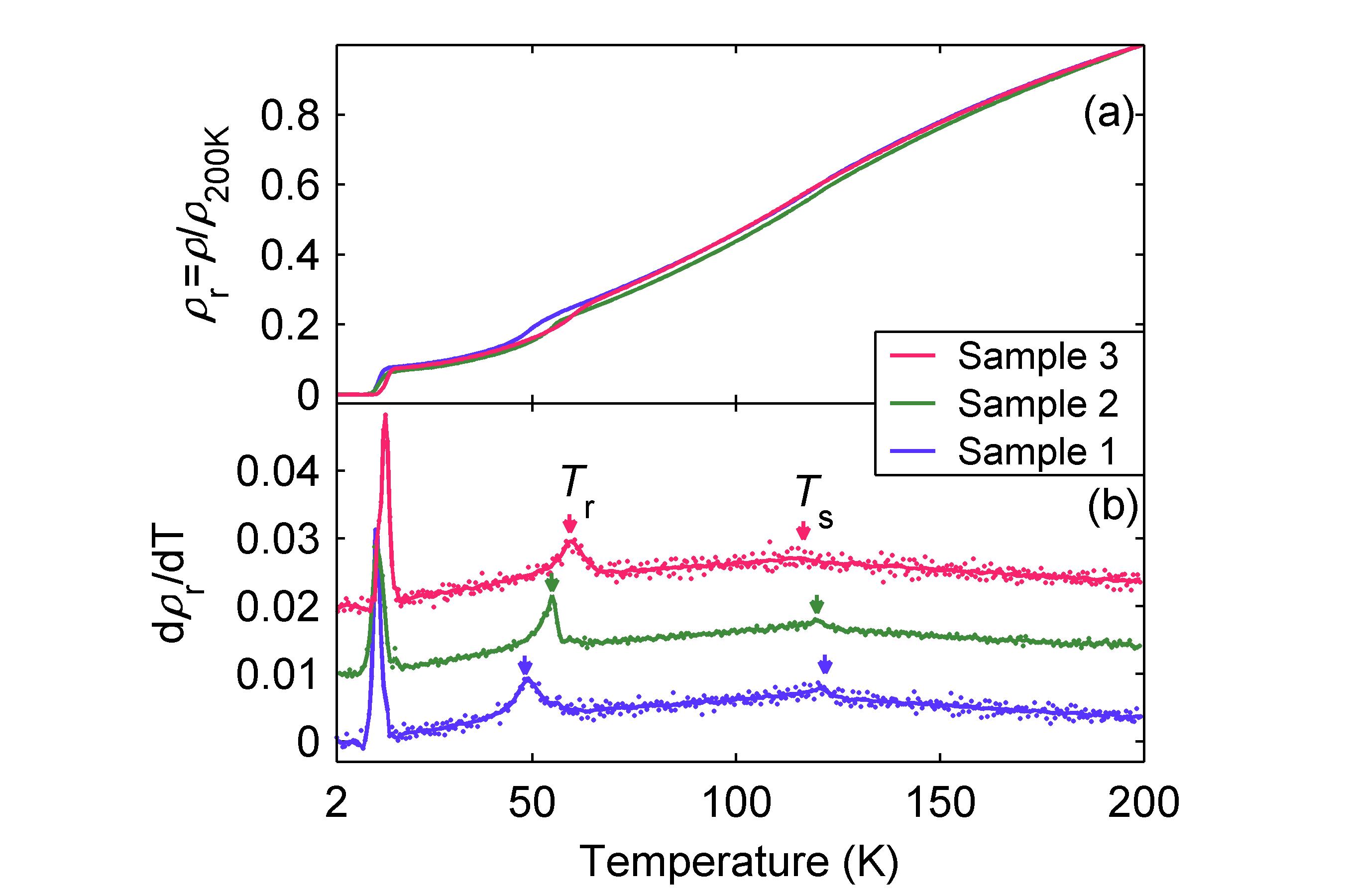}
\caption{\label{Fig1}
$T$ dependence of (a) resistivity and (b) its derivative measured on three samples.}
\label{Fig1}
\end{figure}

The single crystals of Sr$_{1-x}$Na$_x$Fe$_2$As$_2$ used in this study were grown by a self-flux method \cite{TaddeiPRB2016}. Compared to Ba$_{1-x}$Na$_x$Fe$_2$As$_2$ \cite{AvciNatCommu2014} and Ba$_{1-x}$K$_x$Fe$_2$As$_2$ \cite{AllredPRB2015}, Sr$_{1-x}$Na$_x$Fe$_2$As$_2$ exhibits a more robust $t$-SDW phase that spans over wider ranges in both temperature $T$ and hole doping $x$ \cite{TaddeiPRB2016}. To determine the transition temperatures, we measured the $T$-dependent resistivity using a standard four-probe technique with a Quantum Design PPMS.
Figure~\ref{Fig1} displays the resistivity and its $T$-derivative of three samples, where the C$_4$-C$_2$ transition temperature $T_{\mathrm{s}}$ and the reentrant C$_2$-C$_4$ transition temperature $T_{\mathrm{r}}$ can be clearly identified as maxima in the derivative.
According to the phase diagram reported in Ref.~\onlinecite{TaddeiPRB2016}, we estimate $x \approx34\%$ in our samples.
The above three samples, as well as those used in our Raman scattering measurements [Fig.~\ref{Fig2}(a)], came from a plate-like single crystal whose thickness was no more than 300 $\mu$m.
The fact that very thin samples cleaved from the same crystal still exhibit slightly different $T_{\mathrm{s}}$ and $T_{\mathrm{r}}$ indicates the presence of noticeable doping inhomogeneity along the $c$ direction \cite{AllredPRB2015,TaddeiPRB2016}, and we consider the variation of $T_{\mathrm{s}}$ and $T_{\mathrm{r}}$ displayed in Fig.~\ref{Fig1} representative of the uncertainty in our samples.

Our Raman scattering measurements were carried out in a confocal backscattering geometry using a Horiba Jobin Yvon LabRAM HR Evolution spectrometer equipped with a liquid-nitrogen cooled CCD detector. The $\lambda=632.8$ nm line from a He-Ne laser was used for excitation. The laser beam was focused down to a $\sim10 \mu$m-diameter spot on the sample surface, which was held under ultra-high vacuum in a liquid-helium flow cryostat. To reduce heating, we used laser power less than 0.4 mW. Our measurements involved two types of sample surfaces that are parallel to the crystallographic $ab$ and $ac$ planes [insets of Fig.~\ref{Fig2}(a)]. The $ab$ surface was prepared by regular cleaving, whereas the $ac$ surface was obtained by fracturing crystals immediately after freezing in liquid nitrogen \cite{RenPRL2015}. Throughout this work, we use the tetragonal crystallographic notation, where $a$ and $b$ denote the in-plane Fe-As-Fe directions, and $a^\prime$ and $b^\prime$ the Fe-Fe directions 45$^\circ$ from $a$ and $b$.

\begin{figure}[t]
\includegraphics[width=3in]{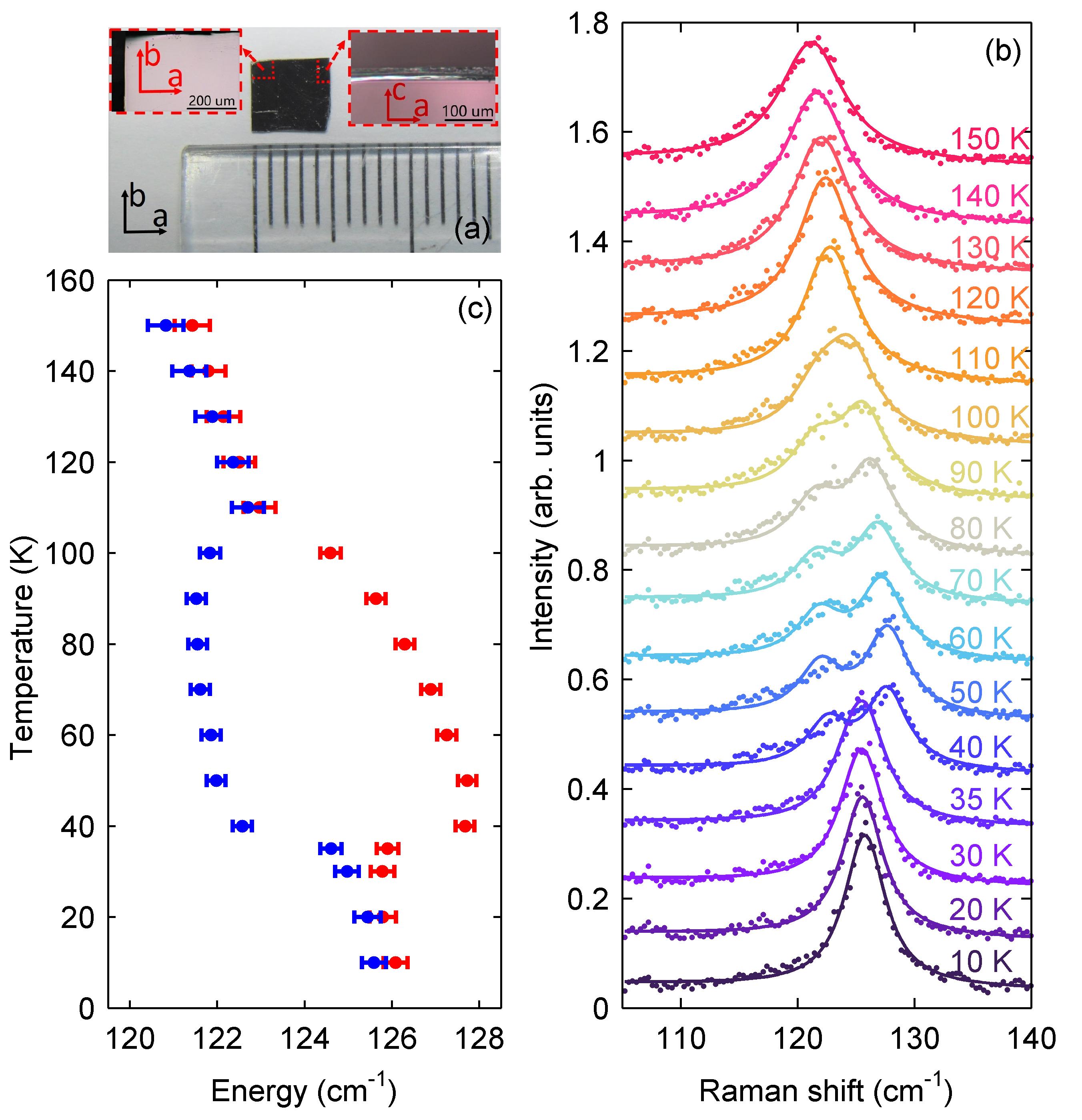}
\caption{\label{Fig2}
(a) Photograph of a single crystal of Sr$_{1-x}$Na$_x$Fe$_2$As$_2$. The insets display freshly-cleaved $ab$ and $ac$ surfaces for our Raman measurements. (b) Raman spectra measured with $ac$ polarizations near the energy of an $E_g$ phonon mode at different temperatures, offset for clarity. (c) $T$ dependence of the fitted peak energies (see text). }
\label{Fig2}
\end{figure}

From factor-group analysis, one expects a total of four Raman-active phonons in the tetragonal phase: $A_{1g}$, $B_{1g}$, and $2\times E_g$. The $E_g$ modes can be detected on the $ac$ surface with $ac$ polarizations of the incident and scattered photons. Because these modes are two-fold degenerate in the tetragonal phase and they become non-degenerate $B_{2g}$ and $B_{3g}$ modes in the orthorhombic phase, the splitting of the corresponding Raman peaks has been previously used to characterize the C$_4$-C$_2$ nematic transition \cite{RenPRL2015,ChauvierePRB2009,HuPRB2016}. As we expect our Sr$_{1-x}$Na$_x$Fe$_2$As$_2$ sample to undergo two transitions related to the breaking and recovery of C$_4$ symmetry, we begin by showing that this is indeed the case from the Raman scattering perspective, and we focus on the low-energy $E_g$ mode around 125 cm$^{-1}$ because its splitting is more prominent. Figure~\ref{Fig2}(b) displays the $T$ evolution of the Raman spectrum. A single peak is observed at 150 K, and upon decreasing $T$ below 110 K, the peak profile substantially changes and becomes more consistent with two peaks. Upon further cooling below $\approx40$ K, however, the two peaks recombine into a single sharp peak.

To quantitatively describe the $T$ evolution, we fit all the spectra in Fig.~\ref{Fig2}(b) with two Lorentzian peaks, the energies of which are displayed in Fig.~\ref{Fig2}(c). When the two fitted peaks are very close together, it is understood that no splitting is observed. It therefore appears that $T_\mathrm{s}$ and $T_\mathrm{r}$ are 100 K-110 K and 30 K-40 K, respectively.
However, the abrupt disappearance of the splitting near $T_\mathrm{r}$, and the likely existence of a total of three peaks in the spectra obtained at 40 K and 50 K [Fig.~\ref{Fig2}(b)], indicate that the C$_2$-C$_4$ transition is of first-order nature \cite{AvciNatCommu2014,AllredPRB2015} with phase coexistence. Moreover, we cannot rule out a distribution of different $T_\mathrm{r}$ in our sample due to doping inhomogeneity. And even though our laser power is low, these measurements on the $ac$ surface (side of thin plate-like crystal) suffer from relatively poor thermal contact between the measured spot and the cold finger where the (nominal) temperature is measured. By further reducing the laser power by 60\%, the splitting is found to disappear at slightly higher temperature, between 40 K and 50 K (not shown), which implies a heating effect of $\Delta T \approx 10$ K in our experiment. After considering this offset and the aforementioned uncertainties, we estimate that $T_\mathrm{s}$ and $T_\mathrm{r}$ are 110 K-120 K and 45 K-60 K, respectively, in good agreement with our resistivity measurements.

\begin{figure}[t]
\includegraphics[width=3.3in]{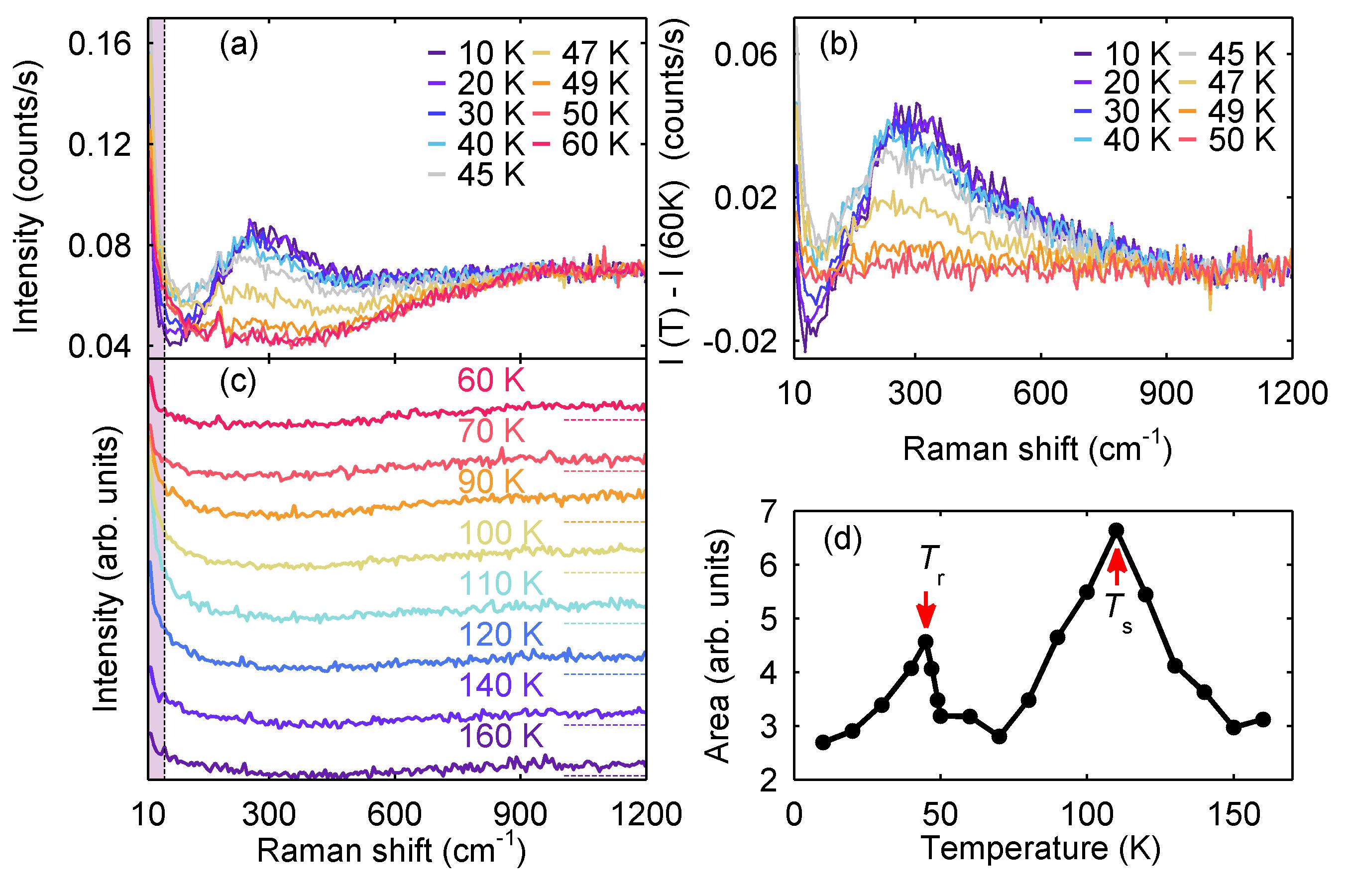}
\caption{\label{Fig3}
Raman spectra measured with $ab$ polarizations. (a) Spectra obtained between 10 K and 60 K. (b) Intensity difference against the spectrum obtained at 60 K. (c) Spectra obtained between 60 K and 160 K, offset for clarity. (d) $T$ dependence of integrated quasi-elastic scattering below intensity 50 cm$^{-1}$ [shaded region in (a) and (c)].}
\label{Fig3}
\end{figure}

Now we turn to measurements of electronic Raman signals, in order to reveal possible critical behaviors as well as characteristic energy scales associated with the C$_4$-C$_2$-C$_4$ transitions. It has been well-established that electronic nematic fluctuations \cite{GallaisPRL2013,KretzschmarNatPhys2016} can be detected in the $B_{2g}$ symmetry channel that is best measured with $ab$ polarizations of the incident and scattered photons. Figure~\ref{Fig3}(a) displays our spectra obtained at low temperatures. A profound broad peak is observed in the $t$-SDW phase around 260 cm$^{-1}$, with a tail that extends up to about 900 cm$^{-1}$. In order to highlight the change upon cooling into the $t$-SDW phase, we plot in Fig.~\ref{Fig3}(b) the intensity difference against the spectrum obtained at 60~K which is just above the transition. Because $ab$ surface has better thermal contact than $ac$ surface, we expect the C$_2$-C$_4$ transition in such measurements to occur at slightly higher (nominal) temperature than that in Fig.~\ref{Fig2}. Indeed, the broad peak becomes noticeable just below 50 K and grows rapidly with decreasing $T$, reaching nearly its full intensity already at 40 K. A close inspection of the data indicates that further development of this peak is accompanied by spectral depletion at lower energies: below 40 K, the low-energy spectral weight is suppressed and redistributed into the peak's high-energy shoulder.

As we already know from previous reports \cite{AvciNatCommu2014,AllredPRB2015} and from our own measurement of the $E_g$ phonon (Fig.~\ref{Fig2}), the transition into the $t$-SDW phase is strongly first-order. Since phase coexistence and random nucleation are generically expected at first-order transitions, it is possible that the sample effectively becomes more disordered near $T_\mathrm{r}$, which might result in additional quasi-elastic scattering of photons. To address this possibility, we analyze the quasi-elastic scattering intensity in our data below 50 cm$^{-1}$ as a function of temperature, and for completeness we treat the data obtained above 60 K [Fig.~\ref{Fig3}(c)] in the same fashion. Consistent with previous reports \cite{GallaisPRL2013,KretzschmarNatPhys2016}, additional scattering signals are observed near $T_\mathrm{s}$ as a result of critical fluctuations associated with the second-order nematic transition. Near $T_\mathrm{r}$, we indeed observe a distinct increase in the scattering signal, but different from the critical scattering near $T_\mathrm{s}$, the increase with cooling towards $T_\mathrm{r}$ is more abrupt, and because the transition at $T_\mathrm{r}$ is \textit{not} second-order, it cannot be due to critical scattering. We attribute this intensity increase to diffuse quasi-elastic photon scattering caused by disorders, and/or by quantum tunneling between the closely competing $o$-SDW and $t$-SDW states. With this in mind, we believe that the aforementioned rapid development of the broad peak between 50 K and 40 K is also accompanied by low-energy spectral weight depletions, but such depletions are difficult to observe because of the concurrent increase of the quasi-elastic scattering.

The pile-up of intensity near 260 cm$^{-1}$ and above, along with intensity depletions at lower energies, indicates the opening of a new electronic gap in the $t$-SDW phase. Indeed, gap-opening behaviors associated with the $o$-SDW phase have been consistently observed in both optical conductivity \cite{HuPRL2008} and Raman scattering measurements \cite{SugaiJSupNovMagn2011,ChauvierePRB2011} in the ``122'' parent compound BaFe$_2$As$_2$.
In both cases two characteristic energies are observed, which manifest themselves as peak-like features at about 360 cm$^{-1}$ and 890 cm$^{-1}$ in the optical conductivity data \cite{HuPRL2008}.
In the Raman data \cite{SugaiJSupNovMagn2011,ChauvierePRB2011}, the lower characteristic energy manifests itself as a step-like anomaly at about 400 cm$^{-1}$ in \textit{all} symmetry channels, whereas the higher characteristic energy is only observed in the $B_{2g}$ symmetry channel as a broad peak at about 900 cm$^{-1}$.
All of these features become less pronounced and they move to lower energies with electron doping \cite{ChauvierePRB2011,NakajimaPRB2010}, presumably because of increasing mismatch between the hole and electron pockets \cite{LiuNatPhys2010,ZabolotnyyNature2009}.

With commonalities in the large energy width, the $B_{2g}$ symmetry, and the temperature dependence, our result in Fig.~\ref{Fig3} resembles the high-energy Raman peak in the $o$-SDW phase of BaFe$_2$As$_2$ \cite{SugaiJSupNovMagn2011,ChauvierePRB2011}, further supporting the notion that it originates from a new electronic gap in the $t$-SDW phase. The fact that we do not observe any peak-like features in the $o$-SDW phase implies that the mismatch between the hole and electron pockets is already substantial at our doping level, and that the formation of the $t$-SDW phase requires a different type of nesting. Nevertheless, the ratio between the $o$-SDW and $t$-SDW transition temperatures in BaFe$_2$As$_2$ and our sample, respectively, $T_\mathrm{s}(137\,\mathrm{K}) / T_\mathrm{r}(50\,\mathrm{K}) = 2.74$, and the ratio between the corresponding broad-peak energies \cite{SugaiJSupNovMagn2011,ChauvierePRB2011}, $(900\,\mathrm{cm}^{-1})/(260\,\mathrm{cm}^{-1})=3.46$, are roughly consistent. Our results are qualitatively consistent with a recent optical conductivity study of Ba$_{1-x}$K$_x$Fe$_2$As$_2$ ($x=0.247$), which exhibits the C$_2$-C$_4$ transition at $T_\mathrm{r}=32$~K \cite{MallettPRL2015}. In the optical data, the high-energy peak associated with the $o$-SDW phase is at lower energy and less pronounced than in the parent compound \cite{HuPRL2008}, but most importantly, additional spectral weight is observed in the $t$-SDW phase at energies below the $o$-SDW high-energy peak. Very recently, electronic band splittings associated with the $t$-SDW phase were detected by photoemission experiments in Ba$_{0.75}$Na$_{0.25}$Fe$_2$As$_2$ \cite{YiCommunication}, with a characteristic energy that is roughly consistent with our results.

\begin{figure}
\includegraphics[width=3.3in]{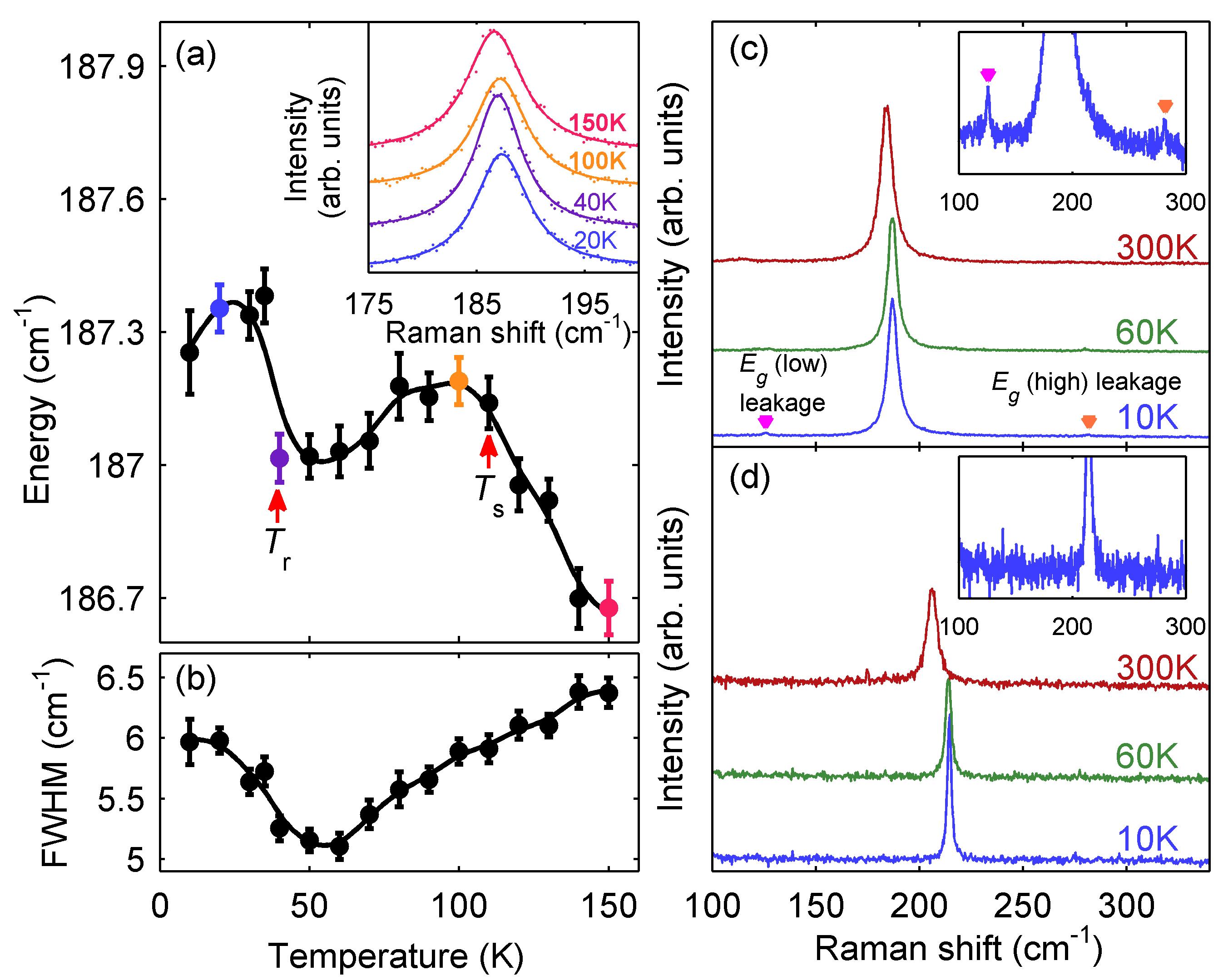}
\caption{\label{Fig4}
$T$ dependence of (a) the energy and (b) the full-width at half-maximum (FWHM) of the $A_{1g}$ phonon peak. The inset of (a) presents four representative spectra to demonstrate the non-monotonic peak movement. (c and d) Raman spectra measured with $cc$ and $aa$ polarizations, respectively, at selected temperatures. The insets are zoom-in view of the 10 K spectra. Arrows in (c) indicate polarization leakage from $E_g$ phonon signals which are unrelated to phonon back-folding.}
\label{Fig4}
\end{figure}

Finally, it is expected that there may be $(\pi, \pi)$ charge order accompanying \cite{ChristensenPRB2015,GastiasoroPRB2015} or even preceding \cite{FernandesPRB2016} the $t$-SDW order. While the in-plane charge order only breaks a glide-plane symmetry, the magnetic and non-magnetic Fe must stack in an aligned fashion \cite{AllredNatPhys2016} along the $c$ direction in order to restore the global C$_4$ symmetry, and this enlarges the structural primitive cell by breaking the body-center translation symmetry. Therefore, we have performed a deliberate search for phonon back-folding behaviors, which are generally expected when the lattice translation symmetry is broken by charge order \cite{HolyPRB1977,LazarevicPRB2011,DuPRB2014,HuPRB2015,AlbertiniPRB2016}. Figure~\ref{Fig4} displays the result of our search with $cc$ and $aa$ polarizations, in which the $A_{1g}$ and $B_{1g}$ phonon signals are most intense, respectively. As the sample undergoes the C$_4$-C$_2$ and then the C$_2$-C$_4$ transitions upon cooling, clear anomalies are observed in both the energy and the line width of the $A_{1g}$ phonon [Figs.~\ref{Fig4}(a-b)], which indicates that the lattice dynamics are indeed affected by the magnetic order. However, we do not observe any additional phonon peaks in the $t$-SDW phase [Figs.~\ref{Fig4}(c-d)], to an accuracy of about 0.3\% of the $A_{1g}$ and 2\% of the $B_{1g}$ peak amplitude, which are the statistical uncertainty of our data after about one hour of accumulation at each temperature. Similar searches have been performed with other polarizations ($ab$, $a^\prime b^\prime$, $a^\prime a^\prime$, and $ac$) but no additional phonon was observed either. The lack of pronounced phonon back-folding behavior suggests that any charge-order-induced translational symmetry breaking must be very weak, and that such charge order cannot be the primary order parameter of the $t$-SDW phase. We note that our result is at variance with a recent report of infrared-active phonon back-folding in the $t$-SDW phase of Ba$_{1-x}$K$_x$Fe$_2$As$_2$ \cite{MallettPRL2015}. Although the compounds are different, the infrared features are rather pronounced, so given the seemingly universal property of the $t$-SDW phase in the hole-doped pnictides \cite{AllredPRB2015,TaddeiPRB2016,TaddeiPRB2017}, we find it difficult to reconcile the previous result with ours. One possibility is that the infrared features are actually related to the electronic feature that we observe at $\approx 260$ cm$^{-1}$, as their energies are indeed very similar.

In summary, we have performed a systematic Raman scattering study of Sr$_{1-x}$Na$_x$Fe$_2$As$_2$ which exhibits two magnetic transitions. The temperature evolution of an $E_g$ phonon confirms the sequential C$_4$-C$_2$-C$_4$ transitions in our sample. Upon entering the $t$-SDW phase, we observe a distinct $B_{2g}$ electronic feature that indicates the development of a new electronic gap of about 260 cm$^{-1}$, or 32 meV. While charge order may accompany the $t$-SDW order and break the lattice translation symmetry, our deliberate search for phonon back-folding behaviors yields a null result, implying that any such symmetry breaking must be weak.

\begin{acknowledgments}

We wish to thank Fa Wang, Ming Yi, and Donghui Lu for discussions. This work is supported by the National Natural Science Foundation of China (Grants Nos. 11374024, 11522429, 11574004, and 91421107) and Ministry of Science and Technology of China (Grants Nos. 2015CB921302, 2013CB921903, and 2016YFA0301003).
\end{acknowledgments}

\bibliography{SrNa_reference}

\end{document}